Conditioning Medicine
www.conditionmed.org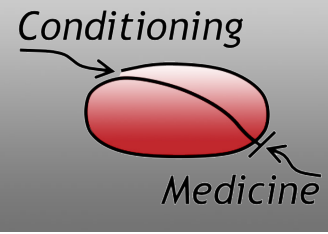

**REVIEW ARTICLE | OPEN ACCESS**

# Requirement for preclinical prioritization of neuroprotective strategies in stroke: Incorporation of preconditioning

Joseph S. Tauskela[1] and Nicolas Blondeau[2][Received: 15 March 2018; accepted: 30 March 2018; published online 28 April, 2018]Acute neuroprotection in numerous human clinical trials has been an abject failure. Major systemic- and procedural-based issues have subsequently been identified in both clinical trials and preclinical animal model experimentation. As well, issues related to the neuroprotective moiety itself have contributed to clinical trial failures, including late delivery, mono-targeting, low potency and poor tolerability. Conditioning (pre- or post-) strategies can potentially address these issues and are therefore gaining increasing attention as approaches to protect the brain from cerebral ischemia. In principle, conditioning can address concerns of timing (preconditioning could be pre-emptively applied in high-risk patients, and post-conditioning after patients experience an unannounced brain infarction) and signaling (multi-modal). However, acute neuroprotection and conditioning strategies face a common translational issue: a myriad of possibilities exist, but with no strategy to select optimal candidates. In this review, we argue that what is required is a neuroprotective framework to identify the "best" agent(s), at the earliest investigational stage possible. This may require switching mindsets from identifying how neuroprotection can be achieved to determining how neuroprotection can fail, for the vast majority of candidates. Understanding the basis for failure can in turn guide supplementary treatment, thereby forming an evidence-based rationale for selecting combinations of therapies. An appropriately designed *in vitro* (neuron culture, brain slices) approach, based on increasing the harshness of the ischemic-like insult, can be useful in identifying the "best" conditioner or acute neuroprotective therapy, as well as how the two modalities can be combined to overcome individual limitations. This would serve as a base from which to launch further investigation into therapies required to protect the neurovascular unit in *in vivo* animal models of cerebral ischemia. Based on these respective approaches, our laboratories suggest that there is merit in examining synaptic activity- and nutraceutical-based preconditioning / acute neuroprotection.[1]National Research Council of Canada, Human Health Therapeutics, Department of Translational Bioscience, 1200 Montreal Road, Ottawa, Ontario, Canada K1A 0R6
[2]Université Côte d'Azur, CNRS, IPMC, UMR7275 Sophia Antipolis, F-06560, France

Correspondence should be addressed to Nicolas Blondeau (blondeau@ipmc.cnrs.fr) and Joseph S. Tauskela (Joseph.Tauskela@nrc-cnrc.gc.ca).**Conditioning Medicine 2018** | Volume 1 | Issue 3 | April 2018

## 1. Introduction

Human clinical trials examining acute administration of a neuroprotective therapeutic after onset of stroke in humans have failed (Labiche and Grotta, 2004). This legacy of failure (>100 trials) has cast a pall on the entire field of neuroprotection in stroke. Business decisions to abandon clinical trials of neuroprotection in stroke (Choi et al., 2014; Wegener and Rujescu, 2013; Howells and Macleod, 2013) have led to a trickle-down effect resulting in diminished commitments in preclinical cerebral ischemia research. With each clinical trial failure, neuroprotection in human stroke is a concept increasingly at risk (Savitz and Fisher, 2007). This is despite the fact that periodic retrospective reviews have identified numerous critical issues in the performance of both human clinical trials and preclinical research. Nonetheless, numerous proposals based on "lessons learned" do not appear to have been sufficient to re-engage the field of neuroprotection in stroke; in fact, the increased rigor and resources now being encouraged to improve performance of studies and trials may actually dissuade increased engagement, in an era of declining resources in this field. Recently developed neuroprotective drugs may also not be viewed as being sufficiently better than those that failed in clinical trials, particularly when a mono-therapy approach continues to dominate trials. It appears the neuroprotection field requires a new strategy. What may be required is tackling the "elephant in the room" – i.e., how therapeutics are chosen at the preclinical level for translation.

## 2. Neuroprotection in Humans

### 2.1 Clinical Trials

Major translational issues identified in human clinical trials are outside the scope of this perspective but, briefly, a common theme noted has been an inability to reproduce several key conditions employed in preclinical animal model studies (Grotta, 2002). The therapeutic dose identified pre-clinically was rarely achieved clinically (if known), due to poor tolerability, safety issues, poor blood-brain barrier permeability or inadequate PK/PD knowledge (Feuerstein et al., 2007). Therapeutic intervention in humans was almost always substantially more delayed from stroke onset than tested in preclinical animal models (or animal models showed that these delays resulted in loss of neuroprotection). Patients were generally not stratified according to stroke severity, compared to the tightly controlled models of cerebral ischemia commonly used in animal models (George and Steinberg, 2015; Minnerup et al., 2014; Kent and Mandava, 2016). Consequently, strokes in some patients were likely more severe than in animal models: all neuroprotective modalities will fail in those brain regions which do not receive timely and adequate reperfusion. Trial durations may have been too short. Importantly, it is generally felt that any one of these deficiencies was severe enough to fully account for the failure of clinical trials (Jonas et al., 1999; Ginsberg, 2008; Zaleska et al., 2009; Jonas et al., 2001). Neuroprotection largely remains in the doldrums at the clinical level, with a few exceptions.

### 2.2 The Legacy

Due to the many issues associated with the performance of human clinical trials, it is difficult to draw conclusions regarding whether the mechanism of action of a therapy contributed to the failure. In principle, understanding the molecular basis for clinical failures can lead to corrective directions for research. However, there has not been much impetus to reverse-translate the conditions in human trials to preclinical animal models, so opportunities to determine individual causes for each failure have been missed (Moskowitz, 2010). In an exception, following the failure of the IMAGES stroke trial on magnesium, a review (Meloni et al., 2006) revealed approximately 50% of preclinical studies reported no neuroprotection, and predicted the failure of the phase III FAST-MAG trial (Meloni et al., 2013).

Considerable uncertainty exists about how therapeutics that failed clinical trials might fare with suggested improvements in procedural failings and advancements in generalization and safety of reperfusion therapy. Hence, one persistent recommendation is to re-investigate drugs which failed in human clinical trials (but were successful pre-clinically), given the use of tPA or increasing adoption of endovascular therapy, coupled with adoption of improvements in better trial designs (Neuhaus et al., 2017; Garber, 2007). However, this approach assumes that more recently developed drugs and therapies would represent no improvement, which seems unlikely. Implicit in this approach is the notion that it is not important (or possible) to rank drugs/therapies according to their neuroprotective potential. And, as discussed below in the preclinical section, the rigor and stringency of preclinical work has generally improved in past efforts.

A concern is that neuroprotection alone may be insufficient if the entire neurovascular unit is not preserved (Lo, 2008; Moskowitz et al., 2010). Neuroprotection might have been achieved in clinical trials (although this is unlikely), but this might have been masked by a failure to protect other components of the neurovascular unit. Hence, the neuro-centric view that has typically prevailed may be destined to fail in principle. However, given the relatively high susceptibility of neurons to ischemia, achievement of neuroprotection must be regarded as a cornerstone, whether achieved by direct or indirect means.

### 2.3 Tackling the Legacy

The widespread abandonment of neuroprotection clinical trials suggests a lack of confidence in any one or more of the requirements to achieve success, including timely application directed at the appropriate target(s) and tissue at a sufficient dose in properly stratified patients. However, some progress is being made. Multimodal imaging studies continue to improve in delineating penumbral regions – or core regions if reperfusion is initiated fast enough – which may be amenable to neuroprotection (Zerna et al., 2016). Reduction in "door-to-needle" times and improved functional outcomes resulting from intravenous thrombosis and endovascular intervention are being made in stroke centers (Saver et al., 2015a). In addition, in the FASTMAG phase 3 trial, three-quarters of patients received treatment within the "golden hour" (the first 60 min from stroke onset) in a pre-hospital setting (Saver et al., 2015b). The ongoing FRONTIER trial evaluates if administration by paramedics in an out-of-hospital setting of NA-1, which uncouples the post-synaptic density protein PSD-95 from excitotoxic signaling pathways (Aarts et al., 2002), reduces disability in stroke patients in Canada (clinicaltrials. gov NCT02315443). The cohort of patients potentially eligible to receive a neuroprotective treatment is expected to increase, based on recent studies suggesting that transient ischemic attacks or minor strokes do indeed possess the capability to induce long-term neurodegeneration (Zamboni et al., 2017; Bivard et al., 2018). Moreover, very recent clinical trials suggest that endovascular treatment may extend the window of intervention (Albers et al., 2018; Nogueira et al., 2018). Thus, the pool of patients potentially eligible to receive neuroprotection may increase considerably.

### 2.4 Strategies in Neuroprotection – What is Missing?

Several approaches have been developed to accelerate a neuroprotective drug/therapy in order to circumvent certain kinds of limitations, including accessibility, regulatory issues, and mono-therapy silos. Accessing the brain by development of BBB-carriers or "shuttles" carrying neuroprotective cargo is becoming closer to reality (Webster and Stanimirovic, 2015;

Stanimirovic et al., 2015). Drug repurposing (or off-labelling) or testing of off-patent drugs has been proposed, in which clinically approved drugs are investigated for neuroprotective potential, thereby bypassing the demanding requirement for regulatory approval. NIH provides access to a library containing >1000 clinically approved drugs, some of which demonstrate neuroprotection (Rothstein et al., 2005; Wang et al., 2006a; Hill et al., 2014). Big pharma has also been granting access to abandoned compounds (Allison, 2012), and other such examples exist. Another approach is mono-therapy which is multi-target in nature, in which one drug exerts an effect at multiple targets in the neurotoxic or NVU toxicity signaling pathways (Lapchak, 2011; Lapchak et al., 2011; Woodruff et al., 2011) but, as an example, albumin failed in a recent clinical trial (Martin et al., 2016; Ginsberg et al., 2013). Hypothermia targets several neurotoxic pathways, as well as the neurovascular unit, and preclinical evidence suggests continued pursuit of this strategy is warranted (van der Worp et al., 2007).

To combat the pervasive mono-therapeutic mentality that has existed throughout the translational pipeline and in regulatory practices for neuroprotection, actual combinations of drugs/therapies should result in more effective targeting of neurotoxic pathways, potentially allowing synergistic neuroprotection and decreased dosing (resulting in fewer adverse effects). Consequently, one tactic which circumvents intellectual property issues and presumably costs is to evaluate a combination of drugs or therapies which are off-patent or not subject to intellectual property concerns. This concept propelled initiation of a small trial using magnesium, Lipitor, minocycline, albumin and hypothermia, in which the authors claimed, "five different areas where you can target the brain – if this does not work, then nothing would work" (Garber, 2007). A subsequent preclinical study examining magnesium sulphate, melatonin and minocycline together did not observe neuroprotection, even with shorter occlusion durations (and therefore smaller infarct volumes) (O'Collins et al., 2011). It cannot be emphasized enough that a risk with any of these approaches – re-investigation of drugs, drug repurposing, a multi-modal drug, or combinations directed exclusively at neurons or the NVU at large – is they may not represent a sufficiently targeted approach, due to a lack of a strong experimentally driven rationale, in which crucial steps in the neurotoxic (and NVU) pathway(s) are being targeted at an appropriate dose with sufficient potency in a temporally relevant manner.

### 3. Neuroprotection in Animal Models
#### 3.1 Preclinical Studies
A major repercussion of the numerous issues identified in human clinical trials is considerable uncertainty as to the true neuroprotective potential of the drugs being investigated, and therefore whether the preclinical animal models of cerebral ischemia are relevant. In fact, the failures in human clinical trials of drugs which showed neuroprotection in animal models of cerebral ischemia have also led to the perception that such models are not predictive or translatable ("everything works in animals but not in humans"). Certainly, limitations and trade-offs with animal models of cerebral ischemia are acknowledged (Sommer, 2017). However, as discussed above, clinical trials generally have not reproduced laboratory conditions, and therefore cannot be used as the basis for concluding that *in vivo* animal models are not predictive.

#### 3.2 The Legacy
It may not even be true that demonstrations of neuroprotection in *in vivo* animal models are sufficiently robust. Numerous issues have been identified in the performance of preclinical studies (Perel et al., 2007; Dirnagl and Endres, 2014; Dirnagl, 2016). The Stroke Therapy Academic Industry Roundtable (STAIR) has been updated several times (1999; Albers et al., 2011), with major recommendations including defining the drug dose-response curve and its time window of efficacy in state-of-the-art models of both permanent and transient occlusions; testing a drug in more than one species, especially non-human primates; integrating co-morbidity into the models (co-morbidity such as hypertension can reduce efficacy of neuroprotectants (Howells and Macleod, 2013)); more evaluation of longer-term functional outcomes; and further incorporation of endovascular treatment (Jovin et al., 2016). Animal models should be chosen to improve mimicry of strokes and physiological conditions in humans (Adkins et al., 2009).

In addition, improved rigor in the performance and reporting of preclinical animal model work is being increasingly encouraged (Dirnagl, 2006; Sena et al., 2007; Howells et al., 2012). As well, studies should be performed in a blinded, unbiased manner, adequately powered to detect a neuroprotective effect, combined with adoption of procedural refinements of *in vivo* experiments (Macleod et al., 2004; Sena et al., 2010; Percie du et al., 2017; Howells and Macleod, 2013). Other major proposals include emulating multicenter stroke trials by preclinical testing of a treatment across several laboratories (Bath et al., 2009; Howells et al., 2012; Dirnagl and Fisher, 2012; Kimmelman et al., 2014; Llovera and Liesz, 2016). A National Institute of Neurological Disorders and Stroke Consensus group recent meeting, resulting in proposals aimed at improving translational stroke research, advocates separation of preclinical studies into two types, with exploratory studies first employed to identify mechanisms of action and high-priority candidates, followed by confirmatory studies in different models and laboratories to build confidence for translation (Bosetti et al., 2017).

#### 3.3 Tackling the Legacy
These calls to action are important, but widespread adoption and implementation may be challenging on a number of levels (Howells and Macleod, 2013). These refinements in preclinical research may be viewed as restrictive or infeasible or, even if implemented, insufficient to improve predictability of how therapeutics might fare in clinical trials. Some approaches require a change in mindset, by convincing laboratories to perform confirmatory studies, which in turn requires acceptance by journals, editors, academia, granting agencies and other parties which can affect publishing, funding and career progression. Moreover, a multi-laboratory preclinical confirmatory approach adhering to best practices (proper statistical analyses, adequately powered studies, double-blinding, randomization and other bias-eliminating approaches) likely represents a greater commitment of resources. Thus, from resource and systemic standpoints, it will be very important to have a sound basis in choosing which neuroprotective modality moves from exploratory to confirmatory studies *in vivo*.

Another pressure being brought to bear on stroke research is its cost, both in absolute and relative terms. The failure of all neuroprotective human clinical trials has resulted in an exodus of stroke research in pharma, academia and government laboratories. This perspective alone places further pressure on candidates entering human clinical trials to succeed, in order to generate the kind of momentum needed to re-energize the research community. Even when optimism was higher two decades ago, funding by granting agencies was far lower for stroke compared to cancer, based on the relative degree of economic impacts exerted on society by these diseases (Rothwell, 2001; Luengo-Fernandez et al., 2015). Of course, advances in neuroprotection in stroke may not derive from strictly targeted work, which makes general funding for basic research important. This too is generally in decline. Consequently, aside from the enormous burden stroke exacts upon society (Flynn et al., 2008), from an economic standpoint of drug discovery, it has become even more important to wisely choose therapeutics at the earliest stage possible in the translational ladder.

*3.4 Strategies in Neuroprotection – What is Missing?*

Substantive neuroprotection may not be achieved even with adopting the numerous recommendations and approaches emanating from clinical human trials and preclinical animal studies. This may be because progress has arguably been insufficient in deciphering how a therapeutic might fare. Simply put, it does not appear that the crisis of confidence facing the neuroprotection field has abated with the current catalog of therapies, nor does it seem this is changing with new developments. The elephant in the room may be how to properly rank therapeutics which have the best chance for translation.

*Having Too Many Choices…*
First, a major issue facing the neuroprotection field is the sheer number of possibilities. Indeed, it is ironic that the acute neuroprotection field is awash in many ostensibly very efficacious preclinical therapies, yet devoid of clinical approval. The preconditioning field (see below) faces a common challenge – and perhaps even more so – with myriad genomic, proteomic and signaling mechanisms having been identified in many different kinds of preconditioning stimuli. (One perplexing issue is that it is not at all clear how such diverse stimuli can result in reductions in infarct injury *in vivo* and neuroprotection *in vitro* (Moskowitz et al., 2010).)

*….Requires Prioritization*
Second, "lessons learned" have been largely focused on improving the process of testing/verifying a therapeutic, but not on how to choose a therapeutic. The challenge of how to prioritize a neuroprotective modality was quantitated over a decade ago in a landmark study by O'Collins et al. (2006), who reported that drugs used in human clinical trials of stroke performed no better than any others in animal model tests of cerebral ischemia. Other research groups report a similar lack of strong basis in preclinical studies for proceeding to clinical trials (van der Worp et al., 2005). Despite this recognition, this situation does not appear to have improved (Philip et al., 2009). Recent entrants into clinical trials may indeed represent improvements based on current understanding of cerebral ischemia and pharmacology, but how would we know? How do we discern which neuroprotective strategy offers the highest therapeutic index, identified at the earliest stage possible, particularly in an era in which performance standards are increasing and engagement is declining? Clearly, it would be advantageous to have a strategy to prioritize neuroprotective treatments and eliminate unworthy candidates, particularly given the high number and variety of candidates. Pursuit of such a mindset may engender increased confidence and re-engage the neuroprotection field.

As important as it is to understand how a therapy will succeed, it is even more important to understand – and indeed, actively pursue – the failure of therapeutics, as early in the translational pipeline as possible. This mantra of "fail early" is usually regarded as a fundamental tenet in the pharmaceutical industry, although clearly not to a great enough extent in the neuroprotection in stroke field. The mantra must considerably expand to include this field as a whole, requiring that societal and mechanistic-based silos be broken down.

## 4. The Case for *In Vitro*

*4.1 Translation From in Vitro to in Vivo to Clinical*

*In vitro* preparations (neuron/astrocyte cultures, brain slices) may be of use in understanding why many neuro-based treatments failed in clinical trials and, with certain modifications, to assist in prioritizing therapeutics for translation. *In vitro* preparations are typically restricted to screening within-family candidates, investigating mechanisms and demonstrating proof-of-concept of neuroprotection, subsequently followed up by an *in vivo* model of ischemia (which generally correlates quite well with i*n vitro* results). Given the demands of *in vivo* studies, and the multitude of candidates among which to choose, it would be quite useful if *in vitro* preparations could also help to prioritize neuroprotective drugs/therapies – as well as combinations – prior to translation to the *in vivo* animal model setting.

Concerns about the reductionist nature of using *in vitro* preparations to model ischemia have generally precluded prioritization exercises, but a body of work suggests stronger consideration may be merited. *In vitro* studies can be quite predictive of neuroprotection *in vivo* and in humans. Some drugs which failed in clinical trials have also failed in *in vitro* studies. For instance, anti-excitotoxic-based *in vitro* studies have shown that rank order of neuroprotection in "stroke in a dish" models in neuron cultures varies as MK-801 > memantine > Mg $\geq$ voltage-gated $Ca^{2+}$ antagonists, correlating with *in vivo* data (Seif el et al., 1990; Gorgulu et al., 2000; Kimura et al., 1998; Koretz et al., 1994; Pringle, 2004). Magnesium represented a major milestone by being administered in a pre-hospital setting, but it nonetheless failed in the FASTMAG trial; potency is a major concern, since this NMDA receptor antagonist loses its neuroprotective properties at depolarized membrane potentials that occur during ischemia *in vivo* and *in vitro*. In both *in vitro* and *in vivo* experiments, neuroprotection by NMDA receptor antagonists correlates with potency, but is lost if administration is delayed too long after the insult (Hartley and Choi, 1989; Tauskela et al., 2016). In a clinical trial, administration of the NA-1 peptide after endovascular repair surgery resulted in a decrease in the number (but not volume) of ischemic infarcts (Hill et al., 2012), and is now in phase 3 clinical trials; this drug first demonstrated neuroprotection *in vitro* (Aarts et al., 2002), which translated to rodent and non-human primate models of cerebral ischemia (Bratane et al., 2011; Cook et al., 2012).

*4.2 Ranking Acute Neuroprotection*

One method which might help identify higher priority candidates is to increase the harshness of an *in vitro* neuron culture (or brain slice) insult, thereby allowing rank orders of neuroprotection to be determined (and, in so doing, allow identification of the most deleterious neurotoxic pathway(s)). For instance, exposing neurons to an insult, such as oxygen-glucose deprivation (OGD), lasting long enough to kill some neurons (lethal insult) to a longer duration capable of killing neurons many times over (supra-lethal insult) has led to the following conclusions: first, increasing the concentration of a drug can increase its neuroprotection, but all drugs reach a plateau with increasing duration of OGD, beyond which they fail; second, each mono-therapy fails at a different duration of OGD, thereby providing a rank order of neuroprotection; and last, combining the strongest neuroprotective agents (targeting different pathways) allows neurons to surpass their individual plateaus and to permit longer durations of OGD to be withstood (Bickler and Hansen, 1994; Lynch, III et al., 1995; Aarts et al., 2003; Bonde et al., 2005; Gwag et al., 1995; Kaku et al., 1993). Overall, the ability of neurons to withstand such extended durations of OGD in the presence of combinations of high concentrations of primarily anti-$Ca^{2+}$ based antagonists reveals an extensive gap between the degree of neuroprotection achievable *in vitro*, versus what is possible *in vivo* (tolerability and accessibility issues preclude this kind of therapy). Thus, the field of neuroprotection may be guilty of considerably under-estimating therapeutic armadas required for neuroprotection against stroke in humans, before even considering issues of patient stratification, re-perfusion and brain delivery.

Ranking neuroprotection candidates can take other forms. For instance, using the example supplied above of testing >1000 clinically approved drugs for re-purposing, many candidates demonstrating neuroprotection during an insult failed if applied after the insult (Wang et al., 2006b; Beraki et al., 2013). Other "stroke in a dish" models might be those intended to be more representative of the extracellular milieu during stroke (Vornov and Coyle, 1991; Rytter et al., 2003; Cronberg et al., 2004).

### 4.3 The Future of In Vitro – Human Neurons and Cells

A concern with *in vivo* (and *in vitro*) models of ischemia is that the mechanism of cell death in rodent neurons or tissues may be different from that in humans. Recent advances in human iPSC-derived neuron cultures – and perhaps eventually 3-dimensional cultures comprised of components of the NVU – should be able to provide considerable insight into this question (Antonic et al., 2012; Holloway and Gavins, 2016). Importantly, recent studies suggest that excitotoxic components of OGD are similar between *in vitro* rodent and human neuron cortical cultures (Gupta et al., 2013; Xu et al., 2016). It will be important to determine if supra-lethal insults result in the same pharmacological profile as observed in rodent neuron cultures.

### 4.4 The Future of In Vitro – Multi-Electrode Arrays

Recent advances in evaluation techniques and throughput of therapeutics tested in *in vitro* neural preparations may be of considerable value in prioritizing therapeutics to be translated for *in vivo* testing. The gold standard in electrophysiological testing is patch-clamp, but this technique requires considerable expertise, is a terminal experiment, and has low throughput (high-throughput patch-clamp has not yet been developed sufficiently for analysis of adherent cells such as neurons). Multi-electrode array (MEA) electrophysiology is making considerable inroads. MEAs allow long-term non-invasive monitoring of electrical activity (spontaneous or evoked) in cultured neurons. MEAs could be very useful in predicting if a drug is tolerable or not by determining the degree to which and nature of how neural electrical activity *in vitro* is altered. This would apply to therapeutics meant to be applied acutely, since numerous clinical trials have failed due to tolerability issues. Also, although preconditioning evokes an endogenous stress response, many agents represent potential neurotoxins. Thus, evaluation by MEAs may provide some valuable insight into the degree of tradeoff required between degrees of protection versus stress response required.

Graph theoretical analyses of cultured neurons have provided considerable insight into network function, in a credible fashion, with traits such as population bursts, neuronal hubs of activity, and avalanches, also identified *in vitro*, *in vivo* and in humans. Recent advances in the number of electrodes (>4000) or use of 48-well MEA plates will allow higher-order network analyses or higher throughput, respectively. MEAs are almost always employed in characterizing human neurons and, by measuring the effect of agonists/antagonists of major excitatory/inhibitory neurotransmitters or transporters on electrical activity, allow a detailed depiction of functional receptors and transporters (on a single electrode or neuron basis if required). Thus, MEAs should represent a considerable advance from imaging or simple live-dead analyses (Frank et al., 2018), to determine if a neuroprotective maneuver – either acute, conditioning or both combined – truly preserves neuronal function and truly retains central hallmarks of network function (Vincent et al., 2013; Tauskela et al., 2008).

## 5. Conditioning

### 5.1 The Concept

Conditioning (pre-, per- or post-) for the purposes of neuroprotection against cerebral ischemia has been of academic interest for some time (Wang et al., 2015), and is now starting to enter the clinical trial arena (Keep et al., 2014). Conceptually, conditioning can in principle address some major mechanistic and procedural challenges associated with acute neuroprotective approaches. By definition, preconditioning is applied pre-emptively, while acute neuroprotection implies treatment soon after onset of cerebral ischemia. The pre-emptive requirement would likely restrict widespread application, but situations can be envisioned in which patients at high risk for stroke within their immediate future may be candidates for preconditioning. Several features of the mechanisms underlying brain self-protection are also observed when conditioning is applied following ischemia (post-conditioning), suggesting that timing may not be a limiting factor in implementation of stimulation of endogenous neuroprotection in a clinical setting. The endogenous nature of the response elicited by preconditioning suggests potentially better tolerability compared to conventional exogenous acute neuroprotection. Preconditioning generally exerts a bimodal mechanism of action: neuroprotective signal transduction pathways are up-regulated and neurotoxic pathways are down-regulated, often in a balanced approach. In contrast, acute neuroprotective drugs usually involve suppressing a specific function, risking interference with normal function (e.g., blocking excitotoxicity with an NMDA receptor antagonist also inhibits glutamatergic signaling in stroke-free regions (Hoyte et al., 2004)).

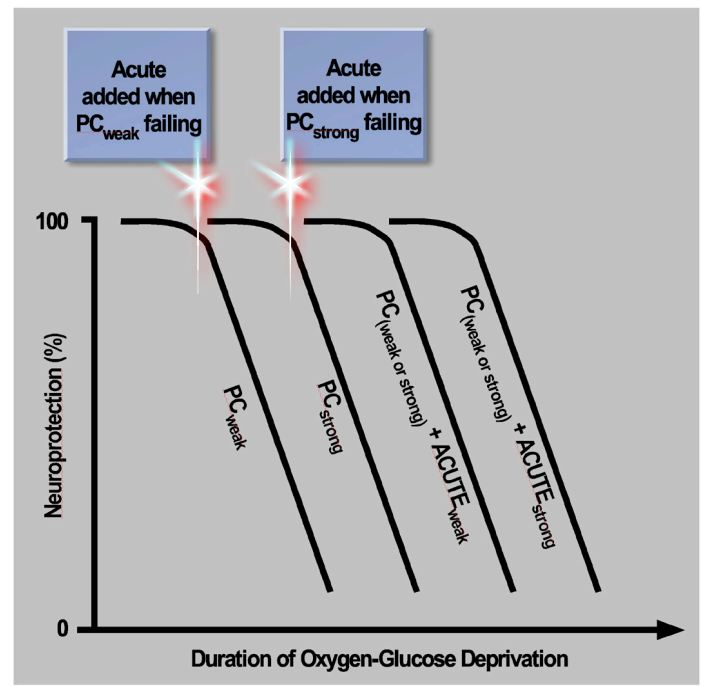

Figure 1. Prioritization of neuroprotective therapeutics and combinations. Subjecting neuron cultures to increasingly longer durations of a stroke-like insult, an oxygen-glucose deprivation (OGD) continuum, can be used to prioritize neuroprotective therapeutics and identify the most efficacious combinations, in an experimentally driven manner. Preconditioning stimuli differ in the ability to protect neurons under severe conditions: strong preconditioners ($PC_{strong}$) protect more neurons at durations of OGD for which weak preconditioners ($PC_{weak}$) are no longer able to protect neurons. At even longer durations, all preconditioning fails to protect neurons, correlating with an inability to suppress a neurotoxic rise in extracellular glutamate levels. Neurons are rescued by adding an NMDA receptor antagonist (labelled $ACUTE_{weak}$) just prior to this rise (i.e., at an earlier time point for neurons subjected to $PC_{weak}$ compared to $PC_{strong}$). Extending the duration of OGD even longer results in loss of neuroprotection again. If, however, a cocktail of anti-$Ca^{2+}$ agents (labelled $ACUTE_{strong}$) is applied at this juncture instead of a single NMDA receptor antagonist, neuroprotection can be restored. Thus, the severity of the insult dictates the nature of the treatment required.

Preconditioning can be applied as one stimulus, but it has a massive polytherapy effect, activating and de-activating numerous genomic, proteomic and signaling pathways. Preconditioning in some instances may be able to circumvent the issue of adequate brain targeting if the brain can be preconditioned in a paracrine manner, therefore not requiring direct access to the brain. For instance, ischemic cuff preconditioning may result in generation and transport of neuroprotective effectors to the brain (Moskowitz and Waeber, 2011). Just as for re-purposing in the acute neuroprotection field, some drugs already in human clinical use have been shown to precondition against cerebral ischemia *in vivo* (Gidday, 2010). As another example of potentially better accessibility, we discuss our experiences with nutraceutical preconditioning below.

Given the backdrop of failure in human neuroprotection, it is appropriate to ask if preconditioning is ready for attempting translation from animal models to humans and, if not, what can be done at the preclinical level to improve the chances of achieving success (Mergenthaler and Dirnagl, 2011). To do otherwise may doom conditioning, resulting in yet another proverbial nail in the coffin of neuroprotection. Conceptually, preconditioning represents a departure from acute neuroprotection, but there are also some important parallels, so some lessons learned with acute neuroprotection may be applied to preconditioning.

### 5.2 Ranking Preconditioning

Borrowing from the acute neuroprotection literature, we have employed supra-lethal OGD in neuron cultures to test a panel of neuronal-targeted preconditioning paradigms – chosen based on a wide range of signaling mechanisms – leading to several conclusions as to how preconditioning might be ranked (Tauskela et al., 2016): (i) A rank order of efficacy was determined for this panel according to the ability of neurons to survive increasingly longer durations of OGD, and a chronic preconditioning stimulus evoking homeostatic downward synaptic scaling was the most neuroprotective. So, to augment Nietzche's quote commonly espoused by the preconditioning community, "That which does not kill us makes us stronger," we would add, "…but may not make us strong enough." (ii) Extending the duration of OGD still further (supra-lethal insult) resulted in loss of neuroprotection by all preconditioners, correlating with an inability to no longer prevent cellular release of glutamate. (iii) In comparison, supra-lethal OGD did not kill cultures co-incubated with MK-801, suggesting superiority over any preconditioning stimulus. (iv) A novel combination therapy was developed based on the presynaptic- and postsynaptic-based mechanisms of preconditioning and MK-801, respectively: specifically, preconditioned cultures could be rescued by timely addition of MK-801 late in the stage of supra-lethal OGD (when extracellular glutamate levels were rising to neurotoxic levels). So, a further extension of Nietzche's quote might be, "That which does not kill us, makes us stronger…but may not make us strong enough…so additional support is required." As indicated above, an NMDA receptor antagonist will also fail if the duration of OGD is further extended, requiring a cocktail of high concentrations of antagonists to provide neuroprotection under these conditions (Fig. 1).

Limited *in vitro* and *in vivo* studies performed elsewhere agree with the concept that preconditioning delays – but does not prevent – damage with prolonged insults, and the concept of ranking. The ischemic preconditioner tested also reached a plateau of neuroprotection if the severity of the ischemic insult was increased (Liu et al., 1992; Shamloo and Wieloch, 1999). Nowak's group reported that preconditioning merely delayed but did not prevent ischemic depolarization in rat (although not in gerbil), which was termed pseudo-preconditioning, since tolerance reflected a decrease in the time the brain was subjected to ischemic depolarization, a key determinant in brain injury (Ueda and Nowak, Jr., 2005). Similarly, chemical preconditioning (3-nitropropionic acid) also delayed depolarization in hippocampal slices (Aketa et al., 2000). Epileptiform preconditioning also failed with increasing severity of cerebral ischemia (Plamondon et al., 1999). Other *in vitro* (Meloni et al., 2002) and *in vivo* (Freiberger et al., 2006) studies limited in scope nonetheless are consistent with different preconditioners providing a relative ranking order.

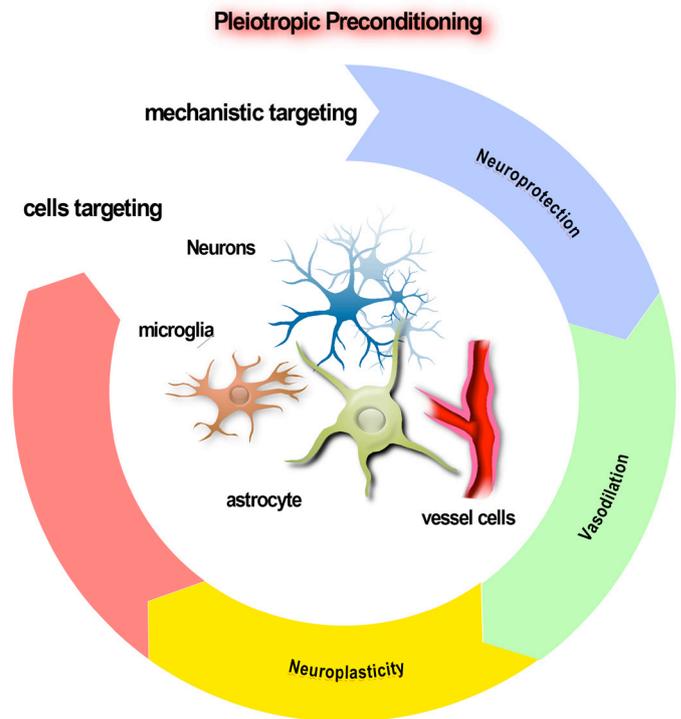

Figure 2. The virtuous circle of pleiotropic conditioning. It will be necessary to develop pleiotropic conditioners possessing additional capabilities beyond directly conferring neuroprotection, in order to enhance survival of the brain after stroke. Targeting of the neurovascular unit (NVU) is desirable, ranging in effects from vasodilation of blood vessels to survival-promotion of glia, setting the stage for enhanced neuroplasticity.

### 5.3 Combination Therapy – Preconditioning "Buys Time" for Acute Drug Intervention

A crucial advantage of preconditioning is therefore to "buy time" before acute pharmacology is needed during the insult, at least *in vitro*. Improvements to each of these two phases should be investigated. For the preconditioning phase, it is not known if combining preconditioners targeting different signaling pathways in neurons – or in astrocytes and glia present in neuron cultures – provides further improvement; i.e., further delaying the time during supra-lethal OGD before acute therapy is required.

For the acute phase, it is necessary to evaluate substituting MK-801 with better-tolerated drugs or therapies, with one possibility being the anti-excitotoxic NA-1 or other peptide-based or small molecule protein-protein inhibitors. Such an approach could consider post-conditioning, and a strategy is required to empirically determine what type of post-ischemic treatment provides the best therapeutic index. Combinations of the acute treatment may need to be considered for longer durations of ischemia. Interestingly, the combination of remote ischemic perconditioning 2 h following embolic middle cerebral artery occlusion in mouse and intravenous tPA at 4 h provided additive neuroprotection (Hoda et al., 2012). Similarly, the combination of pre- or post-conditioning with an acute drug provided better neuroprotection than either alone (McMurtrey and Zuo, 2010). There is a considerable need to build upon this basic neuroprotective combination framework.

*5.4 Alpha-Linolenic Acid (ALA-) Preconditioning*

Due to the efficacy limitations identified in the primarily neuronal-targeted presynaptic-based preconditioners evaluated *in vitro*, it will be necessary to consider pleiotropic conditioners, which possess additional or other capabilities beyond directly conferring neuroprotection, such as displaying abilities to target the neurovascular unit. In such circumstances, these types of preconditioning agents may fail the supra-lethal OGD test *in vitro* but may well yield higher efficacy in *in vitro* and *in vivo* ischemia models adapted to allow as effective a ranking of agents as possible. Apart from neurons, glial or endothelial cells have been successfully preconditioned by many stimuli, so multi-model conditioners able to target the NVU in as comprehensive a manner as possible should be considered (Poinsatte et al., 2015; Stowe et al., 2011). The vasculature is recognized as a key target to truly protect brain (Lo, 2008). There is increasing evidence that preconditioning stimuli, in addition to improving neuronal resistance to stroke, also improve cerebrovascular function (Gidday, 2006) (Fig. 2).

ALA is an omega-3 fatty acid contained in plant-derived edible products. Our research suggests ALA is representative of a direction to follow in order to address several of the deficiencies outlined above. ALA represents a combination of a pleiotropic preconditioner as well as neuroprotective and neuro-restorative features, reviewed in Blondeau (2016). ALA preconditioning induces NFκB, HSP70, BDNF, SNAREs and V-GLUTs, inducing tolerance against two models of neuronal death by excitotoxicity induced by kainic acid injection and global ischemia. ALA preconditioning promotes robust basilar artery dilation, and therefore increases residual cerebral blood flow circulation during ischemia. Intravenous injections of ALA were neuroprotective in animal models ranging from spinal cord injury (Lang-Lazdunski et al., 2003; King et al., 2006; Michael-Titus, 2007) to brain ischemia (Lauritzen et al., 2000; Blondeau et al., 2001; Lang-Lazdunski et al., 2003; Heurteaux et al., 2004; Heurteaux et al., 2006; Blondeau et al., 2002; Blondeau et al., 2007). Moreover, sequential injections of ALA given in the days following ischemia – a protocol could be assimilated to post-treatment as well as post-conditioning – promote long-term survival. Finally, a key component of this form of conditioning is how it is administered: the benefit of preconditioning could be achieved through dietary supplementation, since oral supplementation with ALA reproduced the benefit of ALA preconditioning achieved by injection by also reducing mortality rate and ischemic lesion size (Nguemeni et al., 2010), as well as improving motor and cognitive recovery after 30 min of temporary focal cerebral ischemia (Bourourou et al., 2016). Put in perspective with epidemiological results demonstrating that ALA intake is associated with a lower risk of stroke, the selection of such a natural substance that can neuroprotect the brain in a preconditioning and acute manner suggests that a nutraceutical-based approach represents a paradigm shift in the management of ischemic stroke to circumvent administration and timing issues of traditional drug approaches.

## 6. Conclusions

The field of neuroprotection in stroke is at a crossroads. The legacy of failure in human clinical trials has resulted in a seemingly unassailable challenge permeating to preclinical levels of investigation. A number of practical criteria are converging to a point which now requires tackling how a therapeutic is prioritized within the translational pipeline. Solutions have been offered to improve the rigor of preclinical animal model work but, in an era of ever-shrinking resources, this makes the choice of a therapeutic to be tested in preclinical animal models more important than ever. A myriad of treatment possibilities exists among conditioning and acute neuroprotection, many targeting a seemingly endless array of neurotoxic mechanisms. A framework of experimentally driven prioritization is required at the earliest stage of investigation possible, prior to entry into *in vivo* animal model work. In essence, success can be achieved by actively pursuing failure. In an already extremely difficult field, pursuing this path can be fraught with challenges, both within and outside the laboratory: it requires overcoming a silo mentality which is mechanistic in nature (since combination therapy will probably represent the best chance of achieving substantive neuroprotection), and requires cross-pollination among the many different kinds of societal-based contributors to the neuroprotection field. No longer should investigators be restricted to their particular sub-field or vantage point. We have suggested that proof-in-principle of the kind of prioritization required can be first gained by adjusting *in vitro* models to fail most approaches. We encourage the neuroprotection field to adopt a mindset of seeking prioritization to increase predictive value in translation. This approach of focusing on neuroprotection first at the *in vitro* level is not to exclude therapies adept in targeting the neurovascular unit *in vivo*. This can be a parallel process, and we have highlighted preconditioning, particularly by ALA, since this moiety also possesses acute effects on the NVU.

## References

Stroke Therapy Academic Industry Roundtable (STAIR) (1999) Recommendations for Standards Regarding Preclinical Neuroprotective and Restorative Drug Development. Stroke 30:2752-2758.

Aarts M, Iihara K, Wei W L, Xiong Z G, Arundine M, Cerwinski W, MacDonald J F and Tymianski M (2003) A Key Role for TRPM7 Channels in Anoxic Neuronal Death. Cell 115:863-877.

Aarts M, Liu Y, Liu L, Besshoh S, Arundine M, Gurd J W, Wang Y T, Salter M W and Tymianski M (2002) Treatment of Ischemic Brain Damage by Perturbing NMDA Receptor-PSD-95 Protein Interactions. Science 298:846-850.

Adkins DL, Schallert T and Goldstein L B (2009) Poststroke Treatment: Lost in Translation. Stroke 40:8-9.

Aketa S, Nakase H, Kamada Y, Hiramatsu K and Sakaki T (2000) Chemical Preconditioning With 3-Nitropropionic Acid in Gerbil Hippocampal Slices: Therapeutic Window and the Participation of Adenosine Receptor. Exp Neurol 166:385-391.

Albers GW, Goldstein L B, Hess D C, Wechsler L R, Furie K L, Gorelick P B, Hurn P, Liebeskind D S, Nogueira R G and Saver J L (2011) Stroke Treatment Academic Industry Roundtable (STAIR) Recommendations for Maximizing the Use of Intravenous Thrombolytics and Expanding Treatment Options With Intra-Arterial and Neuroprotective Therapies. Stroke 42:2645-2650.

Albers GW, Marks M P, Kemp S, Christensen S, Tsai J P, Ortega-Gutierrez S, McTaggart R A, Torbey M T, Kim-Tenser M, Leslie-Mazwi T, Sarraj A, Kasner S E, Ansari S A, Yeatts S D, Hamilton S, Mlynash M, Heit J J, Zaharchuk G, Kim S, Carrozzella J, Palesch Y Y, Demchuk A M, Bammer R, Lavori P W, Broderick J P and Lansberg M G (2018) Thrombectomy for Stroke at 6 to 16 Hours With Selection by Perfusion Imaging. N Engl J Med 378:708-718.

Allison M (2012) NCATS Launches Drug Repurposing Program. Nat Biotechnol 30:571-572.

Antonic A, Sena E S, Donnan G A and Howells D W (2012) Human in Vitro Models of Ischaemic Stroke: a Test Bed for Translation. Transl Stroke Res 3:306-309.

Bath PM, Macleod M R and Green A R (2009) Emulating Multicentre Clinical Stroke Trials: a New Paradigm for Studying Novel Interventions in Experimental Models of Stroke. Int J Stroke 4:471-479.

Beraki S, Litrus L, Soriano L, Monbureau M, To L K, Braithwaite S P, Nikolich K, Urfer R, Oksenberg D and Shamloo M (2013) A Pharmacological Screening Approach for Discovery of Neuroprotective Compounds in Ischemic Stroke. PLoS One 8:e69233.

Bickler PE and Hansen B M (1994) Causes of Calcium Accumulation in Rat Cortical Brain Slices During Hypoxia and Ischemia: Role of Ion Channels and Membrane Damage. Brain Res 665:269-276.

Bivard A, Lillicrap T, Marechal B, Garcia-Esperon C, Holliday E, Krishnamurthy V, Levi C R and Parsons M (2018) Transient Ischemic Attack Results in Delayed Brain Atrophy and Cognitive Decline. Stroke 49:384-390.

Blondeau N (2016) The Nutraceutical Potential of Omega-3 Alpha-Linolenic Acid in Reducing the Consequences of Stroke. Biochimie 120:49-55.

Blondeau N, Petrault O, Manta S, Giordanengo V, Gounon P, Bordet R, Lazdunski M and Heurteaux C (2007) Polyunsaturated Fatty Acids Are Cerebral Vasodilators Via the TREK-1 Potassium Channel. Circ Res 101:176-184.

Blondeau N, Widmann C, Lazdunski M and Heurteaux C (2002) Polyunsaturated Fatty Acids Induce Ischemic and Epileptic Tolerance. Neuroscience 109:231-241.

Blondeau N, Widmann C, Lazdunski M and Heurteaux C (2001) Activation of the Nuclear Factor-KappaB Is a Key Event in Brain Tolerance. J Neurosci 21:4668-4677.

Bonde C, Noraberg J, Noer H and Zimmer J (2005) Ionotropic Glutamate Receptors and Glutamate Transporters Are Involved in Necrotic Neuronal Cell Death Induced by Oxygen-Glucose Deprivation of Hippocampal Slice Cultures. Neuroscience 136:779-794.

Bosetti F, Koenig J I, Ayata C, Back S A, Becker K, Broderick J P, Carmichael S T, Cho S, Cipolla M J, Corbett D, Corriveau R A, Cramer S C, Ferguson A R, Finklestein S P, Ford B D, Furie K L, Hemmen T M, Iadecola C, Jakeman L B, Janis S, Jauch E C, Johnston K C, Kochanek P M, Kohn H, Lo E H, Lyden P D, Mallard C, McCullough L D, McGavern L M, Meschia J F, Moy C S, Perez-Pinzon M A, Ramadan I, Savitz S I, Schwamm L H, Steinberg G K, Stenzel-Poore M P, Tymianski M, Warach S, Wechsler L R, Zhang J H and Koroshetz W (2017) Translational Stroke Research: Vision and Opportunities. Stroke 48:2632-2637.

Bourourou M, Heurteaux C and Blondeau N (2016) Alpha-Linolenic Acid Given As Enteral or Parenteral Nutritional Intervention Against Sensorimotor and Cognitive Deficits in a Mouse Model of Ischemic Stroke. Neuropharmacology 108:60-72.

Bratane BT, Cui H, Cook D J, Bouley J, Tymianski M and Fisher M (2011) Neuroprotection by Freezing Ischemic Penumbra Evolution Without Cerebral Blood Flow Augmentation With a Postsynaptic Density-95 Protein Inhibitor. Stroke 42:3265-3270.

Choi DW, Armitage R, Brady L S, Coetzee T, Fisher W, Hyman S, Pande A, Paul S, Potter W, Roin B and Sherer T (2014) Medicines for the Mind: Policy-Based "Pull" Incentives for Creating Breakthrough CNS Drugs. Neuron 84:554-563.

Cook DJ, Teves L and Tymianski M (2012) Treatment of Stroke With a PSD-95 Inhibitor in the Gyrencephalic Primate Brain. Nature 483:213-217.

Cronberg T, Rytter A, Asztely F, Soder A and Wieloch T (2004) Glucose but Not Lactate in Combination With Acidosis Aggravates Ischemic Neuronal Death in Vitro. Stroke 35:753-757.

Dirnagl U (2016) Thomas Willis Lecture: Is Translational Stroke Research Broken, and If So, How Can We Fix It? Stroke 47:2148-2153.

Dirnagl U (2006) Bench to Bedside: the Quest for Quality in Experimental Stroke Research. J Cereb Blood Flow Metab 26:1465-1478.

Dirnagl U and Endres M (2014) Found in Translation: Preclinical Stroke Research Predicts Human Pathophysiology, Clinical Phenotypes, and Therapeutic Outcomes. Stroke 45:1510-1518.

Dirnagl U and Fisher M (2012) International, Multicenter Randomized Preclinical Trials in Translational Stroke Research: It's Time to Act. J Cereb Blood Flow Metab 32:933-935.

Feuerstein GZ, Zaleska M M, Krams M, Wang X, Day M, Rutkowski J L, Finklestein S P, Pangalos M N, Poole M, Stiles G L, Ruffolo R R and Walsh F L (2007) Missing Steps in the STAIR Case: a Translational Medicine Perspective on the Development of NXY-059 for Treatment of Acute Ischemic Stroke. J Cereb Blood Flow Metab 28(1):217-9.

Flynn RW, MacWalter R S and Doney A S (2008) The Cost of Cerebral Ischaemia. Neuropharmacology 55:250-256.

Frank CL, Brown J P, Wallace K, Wambaugh J F, Shah I and Shafer T J (2018) Defining Toxicological Tipping Points in Neuronal Network Development. Toxicol Appl Pharmacol, In press.

Freiberger JJ, Suliman H B, Sheng H, McAdoo J, Piantadosi C A and Warner D S (2006) A Comparison of Hyperbaric Oxygen Versus Hypoxic Cerebral Preconditioning in Neonatal Rats. Brain Res 1075:213-222.

Garber K (2007) Stroke Treatment--Light at the End of the Tunnel? Nat Biotechnol 25:838-840.

George PM and Steinberg G K (2015) Novel Stroke Therapeutics: Unraveling Stroke Pathophysiology and Its Impact on Clinical Treatments. Neuron 87:297-309.

Gidday JM (2010) Pharmacologic Preconditioning: Translating the Promise. Transl Stroke Res 1:19-30.

Gidday JM (2006) Cerebral Preconditioning and Ischaemic Tolerance. Nat Rev Neurosci 7:437-448.

Ginsberg MD (2008) Neuroprotection for Ischemic Stroke: Past, Present and Future. Neuropharmacology 55:363-389.

Ginsberg MD, Palesch Y Y, Hill M D, Martin R H, Moy C S, Barsan W G, Waldman B D, Tamariz D and Ryckborst K J (2013) High-Dose Albumin Treatment for Acute Ischaemic Stroke (ALIAS) Part 2: a Randomised, Double-Blind, Phase 3, Placebo-Controlled Trial. Lancet Neurol 12:1049-1058.

Gorgulu A, Kins T, Cobanoglu S, Unal F, Izgi N I, Yanik B and Kucuk M (2000) Reduction of Edema and Infarction by Memantine and MK-801 After Focal Cerebral Ischaemia and Reperfusion in Rat. Acta Neurochir (Wien ) 142:1287-1292.

Grotta J (2002) Neuroprotection Is Unlikely to Be Effective in Humans Using Current Trial Designs. Stroke 33:306-307.

Gupta K, Hardingham G E and Chandran S (2013) NMDA Receptor-Dependent Glutamate Excitotoxicity in Human Embryonic Stem Cell-Derived Neurons. Neurosci Lett 543:95-100.

Gwag BJ, Lobner D, Koh J Y, Wie M B and Choi D W (1995) Blockade of Glutamate Receptors Unmasks Neuronal Apoptosis After Oxygen-Glucose Deprivation in Vitro. Neuroscience 68:615-619.

Hartley DM and Choi D W (1989) Delayed Rescue of N-Methyl-D-Aspartate Receptor-Mediated Neuronal Injury in Cortical Culture. J Pharmacol Exp Ther 250:752-758.

Heurteaux C, Guy N, Laigle C, Blondeau N, Duprat F, Mazzuca M, Lang-Lazdunski L, Widmann C, Zanzouri M, Romey G and Lazdunski M (2004) TREK-1, a K+ Channel Involved in Neuroprotection and General Anesthesia. EMBO J 23:2684-2695.

Heurteaux C, Laigle C, Blondeau N, Jarretou G and Lazdunski M (2006) Alpha-Linolenic Acid and Riluzole Treatment Confer Cerebral Protection and Improve Survival After Focal Brain Ischemia. Neuroscience 137:241-251.

Hill JW, Thompson J F, Carter M B, Edwards B S, Sklar L A and Rosenberg G A (2014) Identification of Isoxsuprine Hydrochloride As a Neuroprotectant in Ischemic Stroke Through Cell-Based High-Throughput Screening. PLoS One 9:e96761.

Hill MD, Martin R H, Mikulis D, Wong J H, Silver F L, Terbrugge K G, Milot G, Clark W M, Macdonald R L, Kelly M E, Boulton M, Fleetwood I, McDougall C, Gunnarsson T, Chow M, Lum C, Dodd R, Poublanc J, Krings T, Demchuk A M, Goyal M, Anderson R, Bishop J, Garman D and Tymianski M (2012) Safety and Efficacy of NA-1 in Patients With Iatrogenic Stroke After Endovascular Aneurysm Repair (ENACT): a Phase 2, Randomised, Double-Blind, Placebo-Controlled Trial. Lancet Neurol 11:942-950.

Hoda MN, Siddiqui S, Herberg S, Periyasamy-Thandavan S, Bhatia K, Hafez S S, Johnson M H, Hill W D, Ergul A, Fagan S C and Hess D C (2012) Remote Ischemic Perconditioning Is Effective Alone and in Combination With Intravenous Tissue-Type Plasminogen Activator in Murine Model of Embolic Stroke. Stroke 43:2794-2799.

Holloway PM and Gavins F N (2016) Modeling Ischemic Stroke In Vitro: Status Quo and Future Perspectives. Stroke 47:561-569.

Howells DW and Macleod M R (2013) Evidence-Based Translational Medicine. Stroke 44:1466-1471.

Howells DW, Sena E S, O'Collins V and Macleod M R (2012) Improving the Efficiency of the Development of Drugs for Stroke. Int J Stroke 7:371-377.

Hoyte L, Barber P A, Buchan A M and Hill M D (2004) The Rise and Fall of NMDA Antagonists for Ischemic Stroke. Curr Mol Med 4:131-136.

Jonas S, Aiyagari V, Vieira D and Figueroa M (2001) The Failure of Neuronal Protective Agents Versus the Success of Thrombolysis in the Treatment of Ischemic Stroke. The Predictive Value of Animal Models. Ann N Y Acad Sci 939:257-67.

Jonas S, Ayigari V, Viera D and Waterman P (1999) Neuroprotection Against Cerebral Ischemia. A Review of Animal Studies and Correlation With Human Trial Results. Ann N Y Acad Sci 890:2-3.

Jovin TG, Albers G W and Liebeskind D S (2016) Stroke Treatment Academic Industry Roundtable: The Next Generation of Endovascular Trials. Stroke 47:2656-2665.

Kaku DA, Giffard R G and Choi D W (1993) Neuroprotective Effects of Glutamate Antagonists and Extracellular Acidity. Science 260:1516-1518.

Keep RF, Wang M M, Xiang J, Hua Y and Xi G (2014) Full Steam Ahead With Remote Ischemic Conditioning for Stroke. Transl Stroke Res 5:535-537.

Kent TA and Mandava P (2016) Embracing Biological and Methodological Variance in a New Approach to Pre-Clinical Stroke Testing. Transl Stroke Res 7:274-283.

Kimmelman J, Mogil J S and Dirnagl U (2014) Distinguishing Between Exploratory and Confirmatory Preclinical Research Will Improve Translation. PLoS Biol 12 (5):e1001863.

Kimura M, Sawada K, Miyagawa T, Kuwada M, Katayama K and Nishizawa Y (1998) Role of Glutamate Receptors and Voltage-Dependent Calcium and Sodium Channels in the Extracellular Glutamate/Aspartate Accumulation and Subsequent Neuronal Injury Induced by Oxygen/Glucose Deprivation in Cultured Hippocampal Neurons. J Pharmacol Exp Ther 285:178-185.

King VR, Huang W L, Dyall S C, Curran O E, Priestley J V and Michael-Titus A T (2006) Omega-3 Fatty Acids Improve Recovery, Whereas Omega-6 Fatty Acids Worsen Outcome, After Spinal Cord Injury in the Adult Rat. J Neurosci 26:4672-4680.

Koretz B, von B A, Wang N, Lustig H S and Greenberg D A (1994) Pre- and Post-Synaptic Modulators of Excitatory Neurotransmission: Comparative Effects on Hypoxia/Hypoglycemia in Cortical Cultures. Brain Res 643:334-337.

Labiche LA and Grotta J C (2004) Clinical Trials for Cytoprotection in Stroke. NeuroRx 1:46-70.

Lang-Lazdunski L, Blondeau N, Jarretou G, Lazdunski M and Heurteaux C (2003) Linolenic Acid Prevents Neuronal Cell Death and Paraplegia After Transient Spinal Cord Ischemia in Rats. J Vasc Surg 38:564-575.

Lapchak PA (2011) Emerging Therapies: Pleiotropic Multi-Target Drugs to Treat Stroke Victims. Transl Stroke Res 2:129-135.

Lapchak PA, Schubert D R and Maher P A (2011) De-Risking of Stilbazulenyl Nitrone (STAZN), a Lipophilic Nitrone to Treat Stroke Using a Unique Panel of In Vitro Assays. Transl Stroke Res 2:209-217.

Lauritzen I, Blondeau N, Heurteaux C, Widmann C, Romey G and Lazdunski M (2000) Polyunsaturated Fatty Acids Are Potent Neuroprotectors. EMBO J 19:1784-1793.

Liu Y, Kato H, Nakata N and Kogure K (1992) Protection of Rat Hippocampus Against Ischemic Neuronal Damage by Pretreatment With Sublethal Ischemia. Brain Res 586:121-124.

Llovera G and Liesz A (2016) The Next Step in Translational Research: Lessons Learned From the First Preclinical Randomized Controlled Trial. J Neurochem 139 Suppl 2:271-279.

Lo EH (2008) Experimental Models, Neurovascular Mechanisms and Translational Issues in Stroke Research. Br J Pharmacol 153 Suppl 1:S396-405.

Luengo-Fernandez R, Leal J and Gray A (2015) UK Research Spend in 2008 and 2012: Comparing Stroke, Cancer, Coronary Heart Disease and Dementia. BMJ Open 5:e006648.

Lynch JJ, III, Yu S P, Canzoniero L M, Sensi S L and Choi D W (1995) Sodium Channel Blockers Reduce Oxygen-Glucose Deprivation-Induced Cortical Neuronal Injury When Combined With Glutamate Receptor Antagonists. J Pharmacol Exp Ther 273:554-560.

Macleod MR, O'Collins T, Howells D W and Donnan G A (2004) Pooling of Animal Experimental Data Reveals Influence of Study Design and Publication Bias. Stroke 35:1203-1208.

Martin RH, Yeatts S D, Hill M D, Moy C S, Ginsberg M D and Palesch Y Y (2016) ALIAS (Albumin in Acute Ischemic Stroke) Trials: Analysis of the Combined Data From Parts 1 and 2. Stroke 47:2355-2359.

McMurtrey RJ and Zuo Z (2010) Isoflurane Preconditioning and Postconditioning in Rat Hippocampal Neurons. Brain Res 1358:184-90.

Meloni BP, Cross J L, Brookes L M, Clark V W, Campbell K and Knuckey N W (2013) FAST-Mag Protocol With or Without Mild Hypothermia (35 Degrees C) Does Not Improve Outcome After Permanent MCAO in Rats. Magnes Res 26:67-73.

Meloni BP, Majda B T and Knuckey N W (2002) Evaluation of Preconditioning Treatments to Protect Near-Pure Cortical Neuronal Cultures From in Vitro Ischemia Induced Acute and Delayed Neuronal Death. Brain Res 928:69-75.

Meloni BP, Zhu H and Knuckey N W (2006) Is Magnesium Neuroprotective Following Global and Focal Cerebral Ischaemia? A Review of Published Studies. Magnes Res 19:123-137.

Mergenthaler P and Dirnagl U (2011) Protective Conditioning of the Brain: Expressway or Roadblock? J Physiol 589:4147-4155.

Michael-Titus AT (2007) Omega-3 Fatty Acids and Neurological Injury. Prostaglandins Leukot Essent Fatty Acids 77:295-300.

Minnerup J, Wersching H, Schilling M and Schabitz W R (2014) Analysis of Early Phase and Subsequent Phase III Stroke Studies of Neuroprotectants: Outcomes and Predictors for Success. Exp Transl Stroke Med 6:2-6.

Moskowitz MA (2010) Brain Protection: Maybe Yes, Maybe No. Stroke 41:S85-S86.

Moskowitz MA, Lo E H and Iadecola C (2010) The Science of Stroke: Mechanisms in Search of Treatments. Neuron 67:181-198.

Moskowitz MA and Waeber C (2011) Remote Ischemic Preconditioning: Making the Brain More Tolerant, Safely and Inexpensively. Circulation 123:709-711.

Neuhaus AA, Couch Y, Hadley G and Buchan A M (2017) Neuroprotection in Stroke: the Importance of Collaboration and Reproducibility. Brain 140:2079-2092.

Nguemeni C, Delplanque B, Rovere C, Simon-Rousseau N, Gandin C, Agnani G, Nahon J L, Heurteaux C and Blondeau N (2010) Dietary Supplementation of Alpha-Linolenic Acid in an Enriched Rapeseed Oil Diet Protects From Stroke. Pharmacol Res 61:226-233.

Nogueira RG, Jadhav A P, Haussen D C, Bonafe A, Budzik R F, Bhuva P, Yavagal D R, Ribo M, Cognard C, Hanel R A, Sila C A, Hassan A E, Millan M, Levy E I, Mitchell P, Chen M, English J D, Shah Q A, Silver F L, Pereira V M, Mehta B P, Baxter B W, Abraham M G, Cardona P, Veznedaroglu E, Hellinger F R, Feng L, Kirmani J F, Lopes D K, Jankowitz B T, Frankel M R, Costalat V, Vora N A, Yoo A J, Malik A M, Furlan A J, Rubiera M, Aghaebrahim A, Olivot J M, Tekle W G, Shields R, Graves T, Lewis R J, Smith W S, Liebeskind D S, Saver J L and Jovin T G (2018) Thrombectomy 6 to 24 Hours After Stroke With a Mismatch Between Deficit and Infarct. N Engl J Med 378:11-21.

O'Collins VE, Macleod M R, Cox S F, Van R L, Aleksoska E, Donnan G A and Howells D W (2011) Preclinical Drug Evaluation for Combination Therapy in Acute Stroke Using Systematic Review, Meta-Analysis, and Subsequent Experimental Testing. J Cereb Blood Flow Metab 31:962-975.

O'Collins VE, Macleod M R, Donnan G A, Horky L L, van der Worp B H and Howells D W (2006) 1,026 Experimental Treatments in Acute Stroke. Ann Neurol 59:467-477.

Percie du SN, Alfieri A, Allan S M, Carswell H V, Deuchar G A, Farr T D, Flecknell P, Gallagher L, Gibson C L, Haley M J, Macleod M R, McColl B W, McCabe C, Morancho A, Moon L D, O'Neill M J, Perez d P, I, Planas A, Ragan C I, Rosell A, Roy L A, Ryder K O, Simats A, Sena E S, Sutherland B A, Tricklebank M D, Trueman R C, Whitfield L, Wong R and Macrae I M (2017) The IMPROVE Guidelines (Ischaemia Models: Procedural Refinements Of in Vivo Experiments). J Cereb Blood Flow Metab 37:3488-3517.

Perel P, Roberts I, Sena E, Wheble P, Briscoe C, Sandercock P, Macleod M, Mignini L E, Jayaram P and Khan K S (2007) Comparison of Treatment Effects Between Animal Experiments and Clinical Trials: Systematic Review. BMJ 334:197.

Philip M, Benatar M, Fisher M and Savitz S I (2009) Methodological Quality of Animal Studies of Neuroprotective Agents Currently in Phase II/III Acute Ischemic Stroke Trials. Stroke 40:577-581.

Plamondon H, Blondeau N, Heurteaux C and Lazdunski M (1999) Mutually Protective Actions of Kainic Acid Epileptic Preconditioning and Sublethal Global Ischemia on Hippocampal Neuronal Death: Involvement of Adenosine A1 Receptors and K(ATP) Channels. J Cereb Blood Flow Metab 19:1296-1308.

Poinsatte K, Selvaraj U M, Ortega S B, Plautz E J, Kong X, Gidday J M and Stowe A M (2015) Quantification of Neurovascular Protection Following Repetitive Hypoxic Preconditioning and Transient Middle Cerebral Artery Occlusion in Mice. J Vis Exp (99); e52675.

Pringle AK (2004) In, Out, Shake It All About: Elevation of [Ca2+]i During Acute Cerebral Ischaemia. Cell Calcium 36:235-245.

Rothstein JD, Patel S, Regan M R, Haenggeli C, Huang Y H, Bergles D E, Jin L, Dykes H M, Vidensky S, Chung D S, Toan S V, Bruijn L I, Su Z Z, Gupta P and Fisher P B (2005) Beta-Lactam Antibiotics Offer Neuroprotection by Increasing Glutamate Transporter Expression. Nature 433:73-77.

Rothwell PM (2001) The High Cost of Not Funding Stroke Research: a Comparison With Heart Disease and Cancer. Lancet;357:1612-1616.

Rytter A, Cronberg T, Asztely F, Nemali S and Wieloch T (2003) Mouse Hippocampal Organotypic Tissue Cultures Exposed to in Vitro "Ischemia" Show Selective and Delayed CA1 Damage That Is Aggravated by Glucose. J Cereb Blood Flow Metab 23:23-33.

Saver JL, Goyal M, Bonafe A, Diener H C, Levy E I, Pereira V M, Albers G W, Cognard C, Cohen D J, Hacke W, Jansen O, Jovin T G, Mattle H P, Nogueira R G, Siddiqui A H, Yavagal D R, Baxter B W, Devlin T G, Lopes D K, Reddy V K, du Mesnil de R R, Singer O C and Jahan R (2015a) Stent-Retriever Thrombectomy After Intravenous T-PA Vs. T-PA Alone in Stroke. N Engl J Med 372:2285-2295.

Saver JL, Starkman S, Eckstein M, Stratton S J, Pratt F D, Hamilton S, Conwit R, Liebeskind D S, Sung G, Kramer I, Moreau G, Goldweber R and Sanossian N (2015b) Prehospital Use of Magnesium Sulfate As Neuroprotection in Acute Stroke. N Engl J Med 372:528-536.

Savitz SI and Fisher M (2007) Future of Neuroprotection for Acute Stroke: in the Aftermath of the SAINT Trials. Ann Neurol 61:396-402.

Seif el NM, Peruche B, Rossberg C, Mennel H D and Krieglstein J (1990) Neuroprotective Effect of Memantine Demonstrated in Vivo and in Vitro. Eur J Pharmacol 185:19-24.

Sena E, van der Worp H B, Howells D and Macleod M (2007) How Can We Improve the Pre-Clinical Development of Drugs for Stroke? Trends Neurosci 30:433-439.

Sena ES, van der Worp H B, Bath P M, Howells D W and Macleod M R (2010) Publication Bias in Reports of Animal Stroke Studies Leads to Major Overstatement of Efficacy. PLoS Biol 8:e1000344.

Shamloo M and Wieloch T (1999) Changes in Protein Tyrosine Phosphorylation in the Rat Brain After Cerebral Ischemia in a Model of Ischemic Tolerance. J Cereb Blood Flow Metab 19:173-183.

Sommer CJ (2017) Ischemic Stroke: Experimental Models and Reality. Acta Neuropathol 133:245-261.

Stanimirovic DB, Bani-Yaghoub M, Perkins M and Haqqani A S (2015) Blood-Brain Barrier Models: in Vitro to in Vivo

Translation in Preclinical Development of CNS-Targeting Biotherapeutics. Expert Opin Drug Discov 10:141-155.

Stowe AM, Altay T, Freie A B and Gidday J M (2011) Repetitive Hypoxia Extends Endogenous Neurovascular Protection for Stroke. Ann Neurol 69:975-985.

Tauskela JS, Aylsworth A, Hewitt M, Brunette E and Blondeau N (2016) Failure and Rescue of Preconditioning-Induced Neuroprotection in Severe Stroke-Like Insults. Neuropharmacology 105:533-542.

Tauskela JS, Fang H, Hewitt M, Brunette E, Ahuja T, Thivierge J P, Comas T and Mealing G A (2008) Elevated Synaptic Activity Preconditions Neurons Against an in Vitro Model of Ischemia. J Biol Chem 283:34667-34676.

Ueda M and Nowak T S, Jr. (2005) Protective Preconditioning by Transient Global Ischemia in the Rat: Components of Delayed Injury Progression and Lasting Protection Distinguished by Comparisons of Depolarization Thresholds for Cell Loss at Long Survival Times. J Cereb Blood Flow Metab 25:949-958.

van der Worp HB, de Haan P, Morrema E and Kalkman C J (2005) Methodological Quality of Animal Studies on Neuroprotection in Focal Cerebral Ischaemia. J Neurol 252:1108-1114.

van der Worp HB, Sena E S, Donnan G A, Howells D W and Macleod M R (2007) Hypothermia in Animal Models of Acute Ischaemic Stroke: a Systematic Review and Meta-Analysis. Brain 130:3063-3074.

Vincent K, Tauskela J S, Mealing G A and Thivierge J P (2013) Altered Network Communication Following a Neuroprotective Drug Treatment. PLoS One 8:e54478.

Vornov JJ and Coyle J T (1991) Enhancement of NMDA Receptor-Mediated Neurotoxicity in the Hippocampal Slice by Depolarization and Ischemia. Brain Res 555:99-106.

Wang JK, Portbury S, Thomas M B, Barney S, Ricca D J, Morris D L, Warner D S and Lo D C (2006a) Cardiac Glycosides Provide Neuroprotection Against Ischemic Stroke: Discovery by a Brain Slice-Based Compound Screening Platform. Proc Natl Acad Sci U S A 103:10461-10466.

Wang JK, Portbury S, Thomas M B, Barney S, Ricca D J, Morris D L, Warner D S and Lo D C (2006b) Cardiac Glycosides Provide Neuroprotection Against Ischemic Stroke: Discovery by a Brain Slice-Based Compound Screening Platform. Proc Natl Acad Sci U S A 103:10461-10466.

Wang Y, Reis C, Applegate R, Stier G, Martin R and Zhang J H (2015) Ischemic Conditioning-Induced Endogenous Brain Protection: Applications Pre-, Per- or Post-Stroke. Exp Neurol 272:26-40.

Webster CI and Stanimirovic D B (2015) A Gateway to the Brain: Shuttles for Brain Delivery of Macromolecules. Ther Deliv 6:1321-1324.

Wegener G and Rujescu D (2013) The Current Development of CNS Drug Research. Int J Neuropsychopharmacol 16:1687-1693.

Woodruff TM, Thundyil J, Tang S C, Sobey C G, Taylor S M and Arumugam T V (2011) Pathophysiology, Treatment, and Animal and Cellular Models of Human Ischemic Stroke. Mol Neurodegener 6:11-16.

Xu JC, Fan J, Wang X, Eacker S M, Kam T I, Chen L, Yin X, Zhu J, Chi Z, Jiang H, Chen R, Dawson T M and Dawson V L (2016) Cultured Networks of Excitatory Projection Neurons and Inhibitory Interneurons for Studying Human Cortical Neurotoxicity. Sci Transl Med 8:333ra48.

Zaleska MM, Mercado M L, Chavez J, Feuerstein G Z, Pangalos M N and Wood A (2009) The Development of Stroke Therapeutics: Promising Mechanisms and Translational Challenges. Neuropharmacology 56:329-341.

Zamboni G, Griffanti L, Jenkinson M, Mazzucco S, Li L, Kuker W, Pendlebury S T and Rothwell P M (2017) White Matter Imaging Correlates of Early Cognitive Impairment Detected by the Montreal Cognitive Assessment After Transient Ischemic Attack and Minor Stroke. Stroke 48:1539-1547.

Zerna C, Hegedus J and Hill M D (2016) Evolving Treatments for Acute Ischemic Stroke. Circ Res 118:1425-1442.